\documentclass[aps,prl,twocolumn,floats,superscriptaddress]{revtex4-2}
\usepackage{graphicx}
\usepackage{dcolumn}
\usepackage{bm}
\usepackage{amsmath}
\usepackage{color}
\usepackage{amssymb}
\usepackage{ulem}
\usepackage{xcolor}
\usepackage{subfigure}

\usepackage{amssymb}
\usepackage{latexsym}
\usepackage{epsfig}
\usepackage{psfrag}

\usepackage{amsmath,amsfonts,amssymb,bm,ulem}
\usepackage{color}
\usepackage{float}

\graphicspath{{figs/}}

\newcommand{\be}{\begin{equation}}
\newcommand{\ee}{\end{equation}}
\newcommand{\bea}{\begin{eqnarray}}
\newcommand{\eea}{\end{eqnarray}}
\newcommand{\beq}{\begin{equation}}
\newcommand{\eeq}{\end{equation}}

\begin{document}
\author{Artem Abanov}
\affiliation{Department of Physics,  Texas A\&M University, College Station,  TX 77843, USA}

\author{Andrey Chubukov}
\affiliation{Department of Physics, University of Minnesota, Minneapolis, MN 55455, USA}

\title{Does a pole in the self-energy imply a violation of the Luttinger theorem?}
\begin{abstract}
 Luttinger theorem relates Fermi surface volume  to electron density per spin $n$. It  holds in an arbitrary Fermi liquid but is  widely believed  to break when a fermionic self-energy has a pole. 
  We show that this is not necessarily the case.
 As an example, we consider a fermionic system consisting of two gapped bands, originating from a  self-energy $\Sigma (\omega, \mathbf{k})$ with a pole,
  and a narrow residual quasiparticle band crossing $\omega =0$, with a small residue $Z \ll 1$ 
  (a residual Fermi liquid). We consider Model I, in which $Z$ also renormalizes quasiparticle bandwidth $W$ to $ZW$ and Model II, in which $W$ remains unrenormalized. 
    The Green's function for such a system $G(\omega, \mathbf{k})$ contains two zeros and three poles --- one from the quasiparticle band and two from the gapped bands. 
    Yet, we show that in Model I, the Luttinger-Ward analysis is fully applicable 
     for any $Z >0$, 
    and the results are exactly the same as if the gapped bands did not exist.  For Model II, the Luttinger-Ward analysis recovers once $Z$ exceeds some critical value, which depends on fermionic density.
  \end{abstract}

\maketitle

 \section{Introduction}

\begin{figure}[t]
\includegraphics[width=\columnwidth]{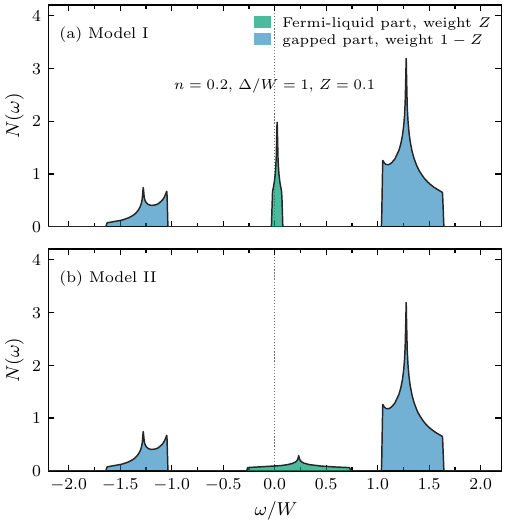}
\caption{\label{fig:DOS} DOS $N(\omega )$  for Model I (a) and  Model II (b)
for $n=0.2$, $\Delta/W=1$ and $Z=0.1$. In both models, the DOS consists of the 
Fermi-liquid part in a patch around $\omega =0$, with total weight $Z$,  and the gapped part with total weight $1-Z$. In between,  the DOS vanishes identically. In Model I, the width near $\omega =0$,  where DOS is non-zero, is compressed by $Z$, whereas in Model II it is not compressed.}
\end{figure}
 
 Luttinger theorem~\cite{Luttinger1960,Luttinger1960a} and its potential violation in systems with either a conventional density-wave order~\cite{Altshuler1997} or a topological order~\cite{Bonetti_2026} have attracted substantial attention from the condensed-matter community in recent years~
 \cite{Georges2001,Dzyaloshinskii2003,Konik2006,*Stanescu2006,Rosch2007,Kiran2013,
 Scheurer2018,*Peralta2023,*Bollmann2024,*Markov2026, Fabrizio2022,*Skolimowski2022,*Blason_2023,*Staffieri_2025, Wagner2023,*Stepanov_2024,*Lehmann_2025,Philips2025}. The theorem states that in an arbitrary Fermi-liquid (FL), even a strongly interacting one, the volume  of the Fermi surface is related to the fermionic density in the same way as in a Fermi gas -- the corrections due to interactions  cancel out.   To prove this, Luttinger and Ward~\cite{Luttinger1960}  and Luttinger~\cite{Luttinger1960a} 
     used the 
 Dyson relation
 between the full Green's function $G(\omega, \mathbf{k})$ and  the single-particle self-energy $\Sigma (\omega, \mathbf{k})$,
 $G^{-1} (\omega, \mathbf{k})  =  \omega -\epsilon_{\mathbf{k}}  - \Sigma (\omega, \mathbf{k})
$ 
in the form 
\begin{equation}\label{eq:LWidentity}
 G(\omega ,\mathbf{k})=\frac{\partial }{\partial \omega }\log G^{-1}(\omega ,\mathbf{k})+G(\omega ,\mathbf{k})\frac{\partial \Sigma (\omega ,\mathbf{k})}{\partial \omega },  
\end{equation}
 and re-expressed the generic relation between the fermionic density $n$  and the frequency integral of the time-ordered Green's function~\cite{Lifshitz1980StatPhys2}: 
\begin{equation}
n = -i \int_{\text{BZ}} \frac{d^{D}k}{(2\pi)^{D}} \lim_{t \to -0}  \int_{-\infty}^\infty  \frac{d\omega}{2\pi}  G (\omega, \mathbf{k}) e^{-i\omega t} ,
 \label{mm_2_1}
\end{equation}
as  
\begin{equation}
n = I_1 + I_2,
 \label{mm_2_2}
\end{equation}
 where 
\begin{equation}
I_1  =  i \int_{\text{BZ}} \frac{d^{D}k}{(2\pi)^{D}}  \lim_{t \to -0}    \int_{-\infty}^\infty \frac{d\omega}{2\pi} \frac{\partial}{\partial \omega} 
\log{ \left[G (\omega,\mathbf{k})\right]} e^{-i\omega t}
 \label{mm_2_3}
 \end{equation}
  and 
  \begin{equation}
I_2  =  -i \int_{\text{BZ}} 
\frac{d^{D}k}{(2\pi)^{D}}   \int_{-\infty}^\infty \frac{d\omega}{2\pi}  G(\omega,\mathbf{k})   \frac{\partial}{\partial \omega} 
\Sigma  (\omega, \mathbf{k}).
 \label{mm_2_4}
 \end{equation}
Here and below, $\mathbf{k}$ is a momentum in a $D$-dimensional Brillouin zone (BZ), and we set the lattice constant to unity, so that the momentum integration is normalized as $\int_{\text{BZ}} \frac{d^{D}k}{(2\pi)^{D}} = 1$.

The condition on time $t$ in Eqs.~(\ref{mm_2_3})--(\ref{mm_2_4})
implies that the contour of integration over $\omega$ has to be extended into the upper half-plane; i.e., only the non-analyticities in the upper half-plane matter.\footnote{There is no need to impose such a condition in $I_2$ as the integral there converges at large frequencies.}  
    
 Luttinger and Ward argued~\cite{Luttinger1960}  that  $I_2 =0$ to any finite order in perturbation theory in the interaction, and that  
 $I_1$ is non-zero due to the non-analyticity of time-ordered $\text{Im}\, G(\omega, \mathbf{k})$ at $\omega \to 0$, where a Fermi liquid (FL) theory is valid, and  is given by  
 \begin{eqnarray}
I_1  &=&  \frac{-i}{2\pi} \int_{\text{BZ}} \frac{d^{D}k}{(2\pi)^{D}}   \log{\left[\frac{i \delta + \epsilon_{\mathbf{k}} -\mu}{-i \delta + \epsilon_{\mathbf{k}} -\mu}\right]} \nonumber \\
 &=& \int_{\text{BZ}} \frac{d^{D}k}{(2\pi)^{D}} \Theta(\mu -\epsilon_{\mathbf{k}}) =\mathcal{V}(\mu)
 \label{mm_2_5}
 \end{eqnarray} 
 where  $\Theta (A) =1$ for $A >0$ and zero otherwise, and  $\mathcal{V}(\mu)$ is the volume bounded by the $(D-1)$-dimensional Fermi surface $\epsilon_{\mathbf{k}}=\mu$.
 A combination of  (\ref{mm_2_2}) and (\ref{mm_2_5}) yields 
 $n = \mathcal{V}(\mu)$, which is the same relation as in a Fermi gas. A more generic formulation  of the  Luttinger theorem has been given in 
~\cite{Oshikawa2000,*Seki2017} (For a comprehensive review of the Luttinger theorem, see Ref.~\cite{Sachdev_book}). Below, we consider, for definiteness, a 2D system
  for which $\mathcal{V}(\mu)$ is the area of the Fermi surface.
 
 Of interest to many researchers in the field is a situation when the system undergoes a phase transition into a state that cannot be adiabatically transformed into a FL~
\cite{Altshuler1997,Georges2001,Dzyaloshinskii2003,Konik2006,*Stanescu2006,Rosch2007,
Sakai2009,Kiran2013,
Fabrizio2022,*Skolimowski2022,*Blason_2023,*Staffieri_2025,Scheurer2018,
*Peralta2023,*Bollmann2024,*Markov2026, Wagner2023,*Stepanov_2024,*Lehmann_2025,Bonetti_2026,Philips2025}. 
 This happens when a fermionic self-energy develops a pole along the real frequency axis. 
  A common perception is that once the self-energy develops a pole,  the original Luttinger-Ward relation between the density $n$ and the Fermi surface area breaks down and is replaced by the new relation for a reconstructed Fermi surface or for an insulator.  A well-known example here is the modification of the Luttinger-Ward relation  in a  state with a charge or spin-density wave order with momentum $\mathbf{Q} = (\pi,\pi)$ (Ref.~\cite{Altshuler1997}). Such a
  state can be viewed as  emerging from the interaction with a boson with a $\delta-$functional propagator $\chi (\Omega,\mathbf{q}) = \delta (\Omega) \delta (\mathbf{q}-\mathbf{Q})$.  This interaction  gives rise to the fermionic self-energy with a pole: 
  $\Sigma (\omega, \mathbf{k}) = \Delta^2/(\omega - (\epsilon_{\mathbf{k}+\mathbf{Q}} - \mu))$, where $\Delta$ is the 
   magnitude of the order. 
  A pole in $\Sigma (\omega, \mathbf{k})$ gives rise to three effects: (i) the Green's function $G(\omega, \mathbf{k})$  vanishes at 
  $\omega = 0$ at 
  $\epsilon_{\mathbf{k}+\mathbf{Q}} =\mu$ (
  sometimes called a "FS" of Green's function zeros), (ii) 
  $G(\omega, \mathbf{k})$ now  has two poles instead of one, and (iii) the frequency derivative of $\Sigma$ has a double pole.   The first two modifications affect $I_1$, which now has contributions from the zero and the two poles; the last 
   modification
   affects $I_2$, which becomes non-zero.  The net result~\cite{Altshuler1997} is the modified Luttinger-Ward relation 
   $S_{h} - S_{e} = 1-n$, where $S_{h}$ and $S_{e}$ 
   are the areas of hole and electron pockets created by the order.     The most relevant result here is the replacement of $n$ by $1-n$. 
   From a physics perspective, this implies that only a fraction of electrons remains gapless and participates in the Luttinger-Ward count. 
   
 In a series of recent publications~\cite{Kiran2013, Gull2013,*Gull2014,
Heath2020,Fabrizio2022,*Skolimowski2022,*Blason_2023,*Staffieri_2025,Scheurer2018,
*Peralta2023,*Bollmann2024,*Markov2026, Wagner2023,*Stepanov_2024,*Lehmann_2025,Bonetti_2026,Philips2025,chubukov_prl_2026,*chubukov_prb_2026}, a self-energy with a pole 
has been suggested 
 for strongly correlated systems that show no signatures of conventional spin or charge order, but a fraction of electrons gets localized. This can happen for several reasons, including  polaron formation in a system with strong electron-phonon interaction~\cite{chubukov_prl_2026,*chubukov_prb_2026} 
    or Mott physics in  systems with Hubbard-type interaction~\cite{Sakai2009,Gull2013,*Gull2014,Fabrizio2022,*Skolimowski2022,*Blason_2023,*Staffieri_2025,Wagner2023,*Stepanov_2024,*Lehmann_2025,Philips2025} 
    (for the latter, a topological order has been suggested as the origin of the pole in $\Sigma (\omega, \mathbf{k})$, Refs.~\cite{Scheurer2018,*Peralta2023,*Bollmann2024,*Markov2026,
    Bonetti_2026}).  It was argued that even in these cases, a pole in $\Sigma (\omega, \mathbf{k})$ leads to the reconstruction of the electronic states and a modification of the Luttinger theorem. 
   
In this communication, we consider an ``intermediate'' situation when the self-energy has a pole with residue $\Delta^2$, and the density of states (DOS) $N(\omega)$  consists of two gapped continua centered at $\omega = \pm \Delta$. At the same time, there 
still  exists a 
 residual FL consisting of quasiparticles with a small residue $Z \ll 1$, which gives rise to 
a non-zero $N(\omega)$ in a narrow region around $\omega =0$. In between this region and the onset of the gapped continua, the DOS vanishes; see Fig.~\ref{fig:DOS}. 
  A DOS in this form 
  was found in the DMFT analysis of the transition between a FL  and a Mott insulator~\cite{Georges1996}. It  also potentially emerges in the polaron state at a non-zero fermionic bandwidth, i.e., away from the atomic limit~\cite{
  freericks1993, Millis_1996,*Millis_1996_a,Ciuchi_1997,*Ciuchi_2003,*Ciuchi_2006,*Ciuchi_2025,Berciu_2006,chubukov_prl_2026,*chubukov_prb_2026}.
 From a physics perspective, we view such a state as one in which some fermions become localized and their DOS is a gapped continuum, while the others  remain in a FL state, albeit with a small quasiparticle residue.

We consider two models.
In the first (Model I), $Z$ only affects the frequency part, like in strong-coupling dynamic theories, its DOS is shown in Fig.~\ref{fig:DOS}(a); in the second (Model II), $Z$ is the overall factor in $G_{FL}$, its DOS is shown in Fig.~\ref{fig:DOS}(b). Notice that in both cases the DOS vanishes in the gaps between the Fermi-liquid part and the gapped part. In Model II, whose Fermi-liquid band keeps the full bandwidth, this requires a larger $\Delta$ than in Model I.  

For both models, the contribution  to the particle number $n$ from the  FL part is
$n_{FL} = Z \int_{\text{BZ}} \frac{d^{D}k}{(2\pi)^{D}} \Theta(\mu -\epsilon_{\mathbf{k}}) = Z 
\mathcal{V}(\mu)$.  We further assume that $N(\omega)$ for the gapped part has $(1-Z)$ as the overall factor, but its shape remains the same as at $Z=0$.  In this approximation, the  contribution  to the particle number $n$ from the  gapped part is $n_{g} = n (1-Z)$. 
Using $n = n_g + n_{FL}$, we find  $\mathcal{V}(\mu) =n$, i.e., the area of the Fermi surface specified by the FL component is equal to the fermionic density, as if the gapped part of $N(\omega)$ did not exist.  
  Our goal is to obtain the values of $I_2$ and $I_1 = n -I_2$ for this case and analyze how they  evolve with $Z$.  For $Z=0$, we show that $I_2$ is non-zero away from half-filling and, moreover, at large $\Delta$, $I_2 =n, I_1 =0$. 
 In  Model II, $I_2$  continuously decreases as $Z$ increases, with the rate set by $\Delta$, 
but in Model I, $I_2$ instantly drops to zero at $Z =0+$  at any $\Delta$, and $I_1$ jumps to become $n$. 
  The outcome here is that the  Luttinger argument that $I_2 =0$, $I_1 =n$  
 remains  valid as long as the FL component is present, despite the
 self-energy having poles and the Green's function having zeros. 

 \section{Model}\label{sec:model}
 
 To be specific, we  consider spinless (spin-polarized)  fermions with tight-binding dispersion $\epsilon_{\mathbf{k}} = -2t (\cos{k_x} + \cos{k_y})$ with bandwidth $W=8t$ at some density $n$ below half-filling ($0<n<1/2$).
   For the model of the gapped continuum, we consider a checkerboard polaron state with momentum $\mathbf{Q} = (\pi,\pi)$ and order parameter $\Delta_Q = \Delta$ 
 (see Refs.~\cite{chubukov_prl_2026,*chubukov_prb_2026} for details). For heavy polarons with energies well above the Debye frequency,
  the corresponding Green's function  $G_{g} (\omega, \mathbf{k})$ can be well approximated by a one-loop form  
    \begin{eqnarray}
    G_{g} (\omega, \mathbf{k}) &=& \left(\omega - (\epsilon_{\mathbf{k}} -\mu_\Delta) - \Delta^2/(\omega + (\epsilon_{\mathbf{k}} - \mu_\Delta))\right)^{-1} \nonumber \\
    &=& \frac{1}{2} \left(\frac{1 - \frac{\epsilon_{\mathbf{k}}-\mu_\Delta}{E_{\mathbf{k}}}}{\omega + E_{\mathbf{k}}} + \frac{1 + \frac{\epsilon_{\mathbf{k}}-\mu_\Delta}{E_{\mathbf{k}}}}{\omega - E_{\mathbf{k}}}\right)   
   \label{1}
     \end{eqnarray}
     where $E_{\mathbf{k}} = \sqrt{\Delta^2 + (\epsilon_{\mathbf{k}} -\mu_\Delta)^2}$.
     This $E$ is finite for any non-zero $\Delta$, i.e., the excitations are fully gapped (the subindex $g$ in $G_g$ stands for gapped).
     In the absence of a FL component, the full
      $G(\omega, \mathbf{k}) = G_{g} (\omega, \mathbf{k})$. 

\begin{figure}[t]
\includegraphics[width=\columnwidth]{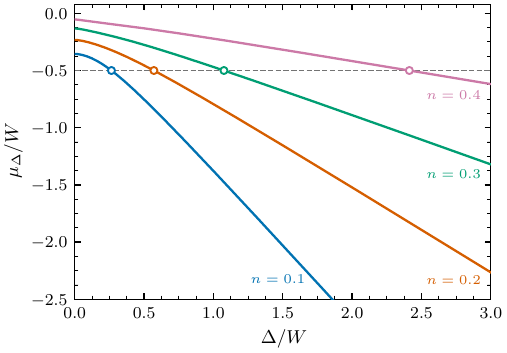}
\caption{\label{fig:muDelta} The chemical potential $\mu_{\Delta }$, 
determined by Eq.~(\ref{2}), as a function of $\Delta/W$ for several filling factors
$n<1/2$.
At $\Delta =0$ we have $\mu_{\Delta }=\mu_{FL}$ (both negative for $n <1/2$); at any non-zero $\Delta$ the electronic spectrum becomes gapped and $\mu_{\Delta }$
moves down. It crosses the bottom of the band, $\mu_{\Delta }=-W/2$ (the dashed line), at
$\Delta =\Delta_c (n)$ (open circles). 
At large $\Delta$,
$\mu_{\Delta } \to -\Delta (1-2n)/(2\sqrt{n(1-n)})$ and 
$\Delta_c (n) \to W\sqrt{n(1-n)}/(1-2n)$. Note that $\Delta_c (n)$ diverges at $n \to 1/2$. }
\end{figure}

      There are two parameters in $G_{g} (\omega,\mathbf{k})$: $\Delta$ and $\mu_\Delta$.  We treat $\Delta$ as a running variable and determine  $\mu_\Delta$ from 
       the  condition on the fermionic density 
\begin{eqnarray}
&&n = \frac{1}{\pi} \int_{\text{BZ}} \frac{d^{2}k}{(2\pi)^{2}}  \int_{-\infty}^0 d\omega  \text{Im}\, G_{g} (\omega,\mathbf{k}) \nonumber \\
&& = \frac{1}{2} \int_{\text{BZ}} \frac{d^{2}k}{(2\pi)^{2}} \left(1 - \frac{\epsilon_{\mathbf{k}}-\mu_\Delta}{E_{\mathbf{k}}}\right)
 \label{2}
     \end{eqnarray} 
We plot $\mu_\Delta$ as a function of $\Delta/W$ for several values of $n<1/2$ in Fig.~\ref{fig:muDelta}.
     
For a FL component, we consider two models:
\begin{itemize}
\item Model I:
 \begin{equation}\label{eq:model1}
G_{FL} (\omega,\mathbf{k}) = \frac{1}{\omega/Z - (\epsilon_{\mathbf{k}} - \mu_{FL})}
.
\end{equation}
This is the case when the fermionic self-energy at small $\omega$ is strong but only depends on frequency. The parameter $Z$ then determines both the quasiparticle residue and the renormalized width of the quasiparticle band $W_{eff} = W Z$.

\item Model II:
\begin{equation}\label{eq:model2}
G_{FL} (\omega,\mathbf{k}) = \frac{Z}{\omega - (\epsilon_{\mathbf{k}} - \mu_{FL})}
.
\end{equation}
This is the case when the self-energy has equally strong frequency and momentum components.
 In this situation, the quasiparticle width is not renormalized. 
 \end{itemize}
 
 The FL contributions to the DOS in the two models are the green areas in panels (a) and (b) in Fig.~\ref{fig:DOS}.  We show below  that the evolution of $I_2$ with $Z$ is discontinuous in Model I and continuous in Model II.
 
For both models, 
\begin{equation}
\frac{1}{\pi} \int_{\text{BZ}} \frac{d^{2}k}{(2\pi)^{2}} \int_{-\infty}^{\infty} d\omega 
\text{Im}\, G_{FL} (\omega,\mathbf{k}) =Z \mathcal{V}(\mu_{FL})
\label{4}
\end{equation}
We take the full Green's function to be
\begin{equation}
G(\omega, \mathbf{k}) = (1-Z) G_{g} (\omega,\mathbf{k}) + G_{FL} (\omega, \mathbf{k})
\label{4a}
\end{equation}
One can easily verify that this  Green's function satisfies the constraint on the total number of states. 

At $\Delta =0$,  $G_{g} (\omega,\mathbf{k}) =  1/(\omega - (\epsilon_{\mathbf{k}} -\mu_{\Delta}))$. Within our scheme, we must reproduce the conventional Fermi gas expression $G (\omega, \mathbf{k}) = 1/(\omega - (\epsilon_{\mathbf{k}} -\mu_{FL}))$ in this limit.  For Model II, this is satisfied  for any $Z$ once we set $\mu_\Delta = \mu_{FL}$; for Model I, this imposes  an additional constraint that $Z$ must be one at $\Delta =0$, i.e., $Z$ and $\Delta$ must be related. The exact relation between $Z$ and $\Delta$ depends on the microscopic details and is not essential for our consideration. 

As we said, we further assume that $G_{g}$ satisfies Eq.~(\ref{2}) even when $Z >0$,
i.e., that the  presence of $G_{FL}$ at $Z >0$ adds the overall factor  $1-Z$ to $G_{g}$, as in 
Eq.~(\ref{4a}), but does not affect its structure.%
\footnote{We emphasize that $\mu_{\Delta }$ and $\mu_{FL}$ are the positions of the common, globally defined chemical potential, measured from the two different band references.}
In this situation, the condition on the total number of particles reduces to 
\begin{equation}
nZ  = \frac{1}{\pi} \int_{-\infty}^{0} d\omega \int_{\text{BZ}} \frac{d^{2}k}{(2\pi)^{2}} \text{Im}\, G_{FL} (\omega,\mathbf{k}) = Z \mathcal{V}(\mu_{FL})
 \label{5}
 \end{equation}
Then, for both Model I and Model II,  
\begin{equation}
\mathcal{V}(\mu_{FL}) = n, 
 \label{5a}
\end{equation}  
i.e.,  the Luttinger-Ward relation between the density and the Fermi surface area is satisfied, as long as $Z >0$.  Our goal is to 
  obtain the value of the Luttinger-Ward integral $I_2$ in this situation and analyze how it varies with $Z$.  

We will see below that in both models, the full Green's function (\ref{4a}) is a {\em rational} function of $\omega + i \delta {\text{sign}} \omega$, with poles and zeros.
 Its logarithmic derivative in the expression for $I_1$ in  (\ref{mm_2_3}) 
 then has only simple poles in the complex $\omega$ plane, which are  located at the positions of the poles and  
 zeros of $G(\omega,\mathbf{k})$ at a given $\mathbf{k}$, with residues $-1$ and $+1$, respectively. By the argument principle,  the frequency integral in (\ref{mm_2_3}) simply counts the number of poles 
  minus the number of zeros of the time-ordered $G (\omega,\mathbf{k})$ in the upper half-plane of complex $\omega$ at a given $\mathbf{k}$, in our case, immediately above the real axis. 
The evaluation of $I_1$ then reduces to locating the poles and the zeros of the Green's function at a given $\mathbf{k}$ with respect to the real axis, and 
\begin{equation}
I_{1}=\int_{\text{BZ}} \frac{d^{2}k}{(2\pi)^{2}}\,I_{1,\mathbf{k}}
,\qquad
I_{1,\mathbf{k}} = \mathcal{P}_{\mathbf{k}} -\mathcal{Z}_{\mathbf{k}}
\label{eq:I1PZ}
\end{equation}
where $\mathcal{P}_{\mathbf{k}}$ is the number of poles of the Green's function, located above the real axis at a given $\mathbf{k}$, and $\mathcal{Z}_{\mathbf{k}}$ is the corresponding number of zeros. 

\section{Case $Z=0$}
\label{sec:Z0}

\begin{figure}[t]
\includegraphics[width=\columnwidth]{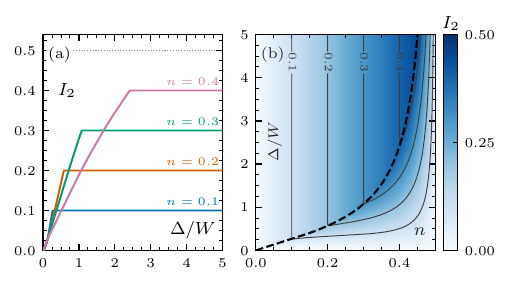}
\caption{\label{fig:I2Z0} The Luttinger-Ward integral $I_{2}$ at $Z=0$, Eq.~(\ref{eq:V:I1I2}).
(a) $I_{2}$ as a function of $\Delta/W$ for several filling factors $n$ in the range $0<n<1/2$.
For any given $n$, $I_2$ initially increases with $\Delta/W$, has a kink at
the  density-dependent $\Delta_c (n)$,
 where $|\mu_\Delta| =W/2$, see Fig. \ref{fig:muDelta}, 
and saturates at larger $\Delta$ at $I_{2}=n$.
The kink is located where $|\mu_{\Delta }|$ reaches the band edge $W/2$ and $\mathcal{V}(\mu_{\Delta })$
vanishes. 
(b) A color-code plot of $I_2 \equiv I_{2}(n,\Delta )$ 
  over the same range of 
  $n$ and $\Delta/W$. The dashed line is $\Delta_c (n)$ and the
thin solid lines are contours of constant $I_{2} (n,\Delta)$.
The value of $\Delta_c (n)$ tends to infinity at $n \to 1/2$, such that in this limit, 
$\Delta/\Delta_c (n \to 1/2) \to 0$  for any finite $\Delta$. As a consequence, 
$I_2 (n \to 1/2, \Delta)$ tends to zero.  } 

\end{figure}

In this case,
 the Fermi-liquid part vanishes, the two models become identical,  and
the time-ordered Green's function is 
\begin{equation}
G(\omega, \mathbf{k}) = G_{g} (\omega,\mathbf{k}) 
 = \frac{\omega+ i\delta {\text{sign}} \omega + \eta_{\mathbf{k}}}{\omega^{2}-E^{2}_{\mathbf{k}} + i \delta},
\label{eq:V:G}
\end{equation}
where we abbreviated
\begin{equation}
\eta_{\mathbf{k}} = \epsilon_{\mathbf{k}} - \mu_{\Delta},
\label{eq:xieta}
\end{equation}
so that $E_{\mathbf{k}} = \sqrt{\Delta^{2}+\eta^{2}_{\mathbf{k}}}$. 

The Green's function (\ref{eq:V:G}) has two poles at $\omega=E_{\mathbf{k}} - i\delta$ and at 
$\omega=-E_{\mathbf{k}} + i\delta$, of which the latter is in the upper half-plane of frequency, and  a single zero at $\omega = -\eta_{\mathbf{k}} + i\delta {\text{sign}} (\eta_{\mathbf{k}})$, which is in the upper half-plane  when $\eta_{\mathbf{k}} > 0$. 
Evaluating the Luttinger-Ward integrals $I_1$ and $I_2$, Eqs.~(\ref{mm_2_3}), (\ref{mm_2_4}),  we obtain
\begin{equation}
I_{1,\mathbf{k}} = \Theta (-\eta_{\mathbf{k}}) 
,\qquad 
I_{1}= \mathcal{V}(\mu_{\Delta}), \qquad 
I_{2} = n - \mathcal{V}(\mu_{\Delta})
\label{eq:V:I1I2}
\end{equation}
where $\mu_{\Delta}$ is determined by (\ref{2}).  At $\Delta =0$, $\mu_{\Delta} = \mu_{FL}$, 
 $\mathcal{V}(\mu_{\Delta}) = n$, and $I_1 =n, I_{2} = 0$, as it should be for free fermions. 
 As $\Delta$ increases at a given density $n$, $\mu_{\Delta}$ increases in magnitude and moves away from $\mu_{FL}$.  We find that in this situation,  
 $I_2$ becomes non-zero and $I_1$ decreases from $n$.  Once  $|\mu_{\Delta}|$ exceeds
 $W/2$, which happens when $\Delta$ exceeds some density-dependent critical $\Delta_c (n)$,
 $\mathcal{V}(\mu_{\Delta})$ vanishes. For larger $\Delta$,  
\begin{equation}
I_1 =0,  \qquad  I_{2} = n.
\label{eq:V:sat}
\end{equation}
 We show this  behavior in Fig.~\ref{fig:I2Z0}. The 
   kinks at different $n$ occur at $\Delta_c (n)$ obtained from (\ref{2}) by setting $\mu_{\Delta } = -W/2$. At the kink, $I_{2}=n$, and it preserves this value for $\Delta > \Delta_c (n)$.
   The dashed line in panel (b) of Fig.~\ref{fig:I2Z0} is the locus of 
   $\Delta_c (n)$. 
 As $n$ approaches half-filling, $\Delta_c (n)$ diverges, and for any given $\Delta$, $I_2 \equiv I_2 (n,\Delta)$  decreases as $n \to 1/2$.  Right at $n=1/2$, $I_2 =0, I_1 =n$ at any finite $\Delta$.
 We discuss the case of half-filling in Sec.~\ref{sec:PH}. 
  
\section{Case $Z > 0$, Model I}\label{sec:modelI}

\begin{figure}[t]
\includegraphics[width=\columnwidth]{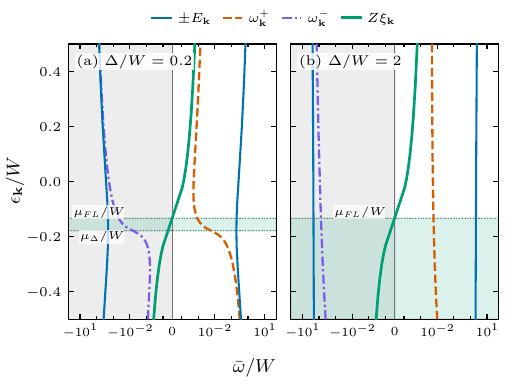}
\caption{\label{fig:pzI} Poles and zeros of the Green's function
$G(\omega, \mathbf{k})$, Eq.~(\ref{eq:VI:G}), in Model I, for $n=0.3$ and $Z=10^{-3}$. Panel (a) -- $\Delta/W=0.2$, where $\mu_{\Delta }$ is still inside the band; panel (b)  -- $\Delta/W=2$,
where $\mu_{\Delta } < -W/2$
 is located below the bottom of the band. The horizontal axis is ${\bar \omega}/W$, where ${\bar \omega} = \omega + i\delta{\text{sign}}\omega$.  
The figure, presented in a symmetric-logarithmic scale, shows how the poles
 (blue and green solid lines, 
${\bar \omega} = \pm E_{\mathbf{k}}$ and 
${\bar \omega} = Z\xi_{\mathbf{k}}$
, respectively)
and  the zeros (orange and purple dashed lines, ${\bar \omega} = \omega^{+}_{\mathbf{k}}$ and ${\bar \omega} = \omega^{-}_{\mathbf{k}}$, respectively) 
  vary as functions of  $\epsilon_{\mathbf{k}}/W$. 
Only poles and zeros at ${\bar \omega} <0$ contribute to the Luttinger-Ward integral $I_1 = 
\int_{\text{BZ}} \frac{d^{2}k}{(2\pi)^{2}} (\mathcal{P}_{\mathbf{k}} -\mathcal{Z}_{\mathbf{k}})$,
where each pole at a given $\mathbf{k}$ contributes $1$ to  $\mathcal{P}_{\mathbf{k}}$ and each zero contributes $1$ to  $\mathcal{Z}_{\mathbf{k}}$. 
Observe that in both panels $\omega^{-}_{\mathbf{k}}$ is negative and $\omega^{+}_{\mathbf{k}}$ is positive for all $\mathbf{k}$. We discuss in the text how these dispersions evolve as $Z \to 0$. }
\end{figure}

 We now keep $Z$ small but finite and include in the full Green's function in (\ref{4a}) 
the Fermi-liquid component (\ref{eq:model1}). 
 We define $\xi_{\mathbf{k}} = \epsilon_{\mathbf{k}} - \mu_{FL}$.  The pole of $G_{FL}$ is at 
  $\omega  = Z\xi_{\mathbf{k}} - i \delta\, {\text{sign}} (\xi_{\mathbf{k}})$ and has a residue $Z$. Combining it with $G_{g}$  into a single expression, we obtain a rational function
\begin{eqnarray}
&&
G(\omega, \mathbf{k}) =
\label{eq:VI:G}\\
&& \frac{{\bar \omega}^{2} + (1-Z)\left(\eta_{\mathbf{k}} - Z\xi_{\mathbf{k}}\right) 
{\bar \omega}
 -  Z\big(E^{2}_{\mathbf{k}} + (1-Z)\,\eta_{\mathbf{k}}\xi_{\mathbf{k}}\big)}
 {\left({\bar \omega} - Z\xi_{\mathbf{k}}\right)\left({\bar \omega}^{2}-E^{2}_{\mathbf{k}}\right)}
.
\nonumber
\end{eqnarray}
 where ${\bar \omega} = \omega + i \delta {\text{sign}} \omega$. 
  This function has three poles and two zeros.  The three poles are at
 $\omega = E_{\mathbf{k}} -i\delta, ~~ \omega =-E_{\mathbf{k}} + i\delta$ and $\omega = Z \xi_{\mathbf{k}} - i \delta{\text{sign}} (\xi_{\mathbf{k}})$. One pole is in the upper half-plane and one is in the lower  half-plane of frequency for all ${\mathbf{k}}$. The third is in the upper half-plane  when 
  $\xi_{\mathbf{k}} <0$, i.e., $\epsilon_{\mathbf{k}} < \mu_{FL}$.
 The number of poles above the real axis is then
 \begin{equation}
 \mathcal{P}_{\mathbf{k}} = 1 + \Theta (-\xi_{\mathbf{k}}) 
\label{eq:VI:P}
\end{equation}
The two zeros ${\omega}^{\pm}_{\mathbf{k}}$ are the roots of the numerator of (\ref{eq:VI:G}). We do not need them separately; to determine their location compared to the real axis, it is enough to know the sign of their product. We have from  (\ref{eq:VI:G})
\begin{equation}
{\omega}^{+}_{\mathbf{k}} {\omega}^{-}_{\mathbf{k}} = - Z f_{\mathbf{k}}
,\qquad
f_{\mathbf{k}} = \Delta^{2} + \eta^{2}_{\mathbf{k}} + (1-Z)\,\eta_{\mathbf{k}}\xi_{\mathbf{k}}
.
\label{eq:VI:prod}
\end{equation}
  We remind that $\eta_{\mathbf{k}} = \epsilon_{\mathbf{k}} - \mu_\Delta$ and $\xi_{\mathbf{k}} = \epsilon_{\mathbf{k}} - \mu_{FL}$. We see that the  function $f_{\mathbf{k}}$  depends on $\mathbf{k}$ only through $\epsilon_{\mathbf{k}}$, and as a function of $\epsilon_{\mathbf{k}}$ is an upward parabola.   We verified that for any $n$, $f_{\mathbf{k}}$ is positive for all ${\mathbf{k}}$ and 
  for all values of $\Delta$ and $Z$, as long as $Z >0$; see Fig.~\ref{fig:pzI}. Then
\begin{equation}
{\omega}^{+}_{\mathbf{k}} {\omega}^{-}_{\mathbf{k}} < 0
,
\label{eq:VI:sign}
\end{equation}
which implies that for any $\mathbf{k}$, one zero out of two is above the real axis, i.e., $\mathcal{Z}_{\mathbf{k}}=1$.
Substituting $\mathcal{P}_{\mathbf{k}}$ and $\mathcal{Z}_{\mathbf{k}}$ into (\ref{eq:I1PZ}), we obtain $I_{1,\mathbf{k}} = \Theta (-\xi_{\mathbf{k}})$ and, after the integration over the BZ, 
\begin{equation}
I_{1} = \mathcal{V}(\mu_{FL}) = n,~~I_2 = n -I_1 =0
\label{eq:VI:I2}
\end{equation}
We emphasize that this holds for any $Z >0$.  We see that the  Luttinger theorem is  satisfied in its original form, for arbitrarily small $Z$ and arbitrarily large $\Delta$,  even though the self-energy has a pole and  the spectral weight chiefly resides in the two gapped continua. The mechanism is transparent: of the two poles in $G$  from the gapped continuum, the one at ${\bar \omega}=+E_{\mathbf{k}}$ never contributes, and the contribution from the other at ${\bar \omega}=-E_{\mathbf{k}}$ 
cancels with the contribution from the Green's function zeros, of which one is above the real axis for all momenta. The only non-zero contribution to $I_1$ then comes from  the pole in
the Fermi-liquid component,  and this contribution is precisely the volume bounded by $\epsilon_{\mathbf{k}} = \mu_{FL}$.

Comparing (\ref{eq:VI:I2}), (\ref{eq:V:sat}) 
with (\ref{eq:V:I1I2}),  we see that $I_1$ and $I_{2}$ are discontinuous at $Z=0+$: 
$I_1$ jumps  from  $I_1 <n$ at $Z=0$ to  $I_1 =n$ at $Z>0+$,  while
$I_2$ drops from  a finite value at $Z=0$ to  $I_2=0$ at $Z = 0+$.
To understand the origin of the jump, we plot in Fig.~\ref{fig:pzI} the positions of the poles and zeros of the Green's function at different $\epsilon_{\mathbf{k}}$ for $n=0.3$ and 
$\Delta/W =0.2$ in panel (a) and $\Delta/W =2$ in panel (b) for a small but finite $Z =10^{-3}$.
 In panel (a),  $\mu_\Delta > -W/2$ is inside the band;  in panel (b), $\mu_\Delta < -W/2$ is below the bottom of the band.  The positions of poles are  blue and green solid lines, and the positions of zeros are the orange and purple dashed lines.  Only poles and zeros at negative $\bar {\omega}$  matter.    Each pole at a given $\mathbf{k}$ contributes $1$ to $\mathcal{P}_{\mathbf{k}}$ and each zero contributes $1$ to $\mathcal{Z}_{\mathbf{k}}$, which enter $I_{1,\mathbf{k}}$ with opposite signs.  In both panels, we see that there is one pole and one zero at ${\bar \omega} <0$ for all $\mathbf{k}$, whose contributions to $I_1$ cancel out, and the contribution from the quasiparticle pole (green line) at $\epsilon_{\mathbf{k}} < \mu_{FL}$. The latter is then the sole contribution to $I_1$, and it yields $I_1 =n$ and hence $I_2=0$.  

 We also extract from  these plots 
   how poles and zeros evolve at  $Z \to 0$. For poles:  the blue line remains intact, while  the green line  becomes a vertical line at ${\bar \omega} =0$. For zeros: in panel (a), the dashed purple line 
   $\omega^{-}_{\mathbf{k}}$ shifts to  $\bar{\omega}=0$ 
   at  $\epsilon_{\mathbf{k}} < \mu_\Delta$ 
   and merges with the green line.
   The value of $\omega^{-}_{\mathbf{k}}$ changes discontinuously at 
   $\epsilon_{\mathbf{k}} < \mu_\Delta$, and at larger $\epsilon_{\mathbf{k}}$
     it is located at ${\bar \omega} <0$, close to the 
     blue line.  Simultaneously, the dashed orange line 
     shifts to ${\bar \omega} =0$ and merges with the green line 
     at $\epsilon_{\mathbf{k}} > \mu_\Delta$.  The poles and zeros that merge along  ${\bar \omega} =0$  just cancel out in the Green's function. As a consequence, at negative ${\bar \omega}$ we are left with the pole (the blue line) at all $\mathbf{k}$ and the zero (the purple dashed line) at $\epsilon_{\mathbf{k}} > \mu_\Delta$. The contributions from poles and zeros then cancel  at $\epsilon_{\mathbf{k}} > \mu_\Delta$, but at $\epsilon_{\mathbf{k}} < \mu_\Delta$ the only contribution is $\mathcal{P}_{\mathbf{k}} =1$ from the blue line.  The net result at $Z=0$ is then  $I_1 = \mathcal{V}(\mu_{\Delta})$, as in  Eq.~(\ref{eq:V:I1I2}). In panel (b), the limiting behavior at $Z =0$ is the same except that now the dashed purple line of zeros remains at ${\bar \omega} <0$ 
      for all $\epsilon_{\mathbf{k}}$, hence $\mathcal{P}_{\mathbf{k}} -\mathcal{Z}_{\mathbf{k}} =0$ for all $\mathbf{k}$. Then $I_1=0$, $I_2 =n$, as in Eq.~(\ref{eq:V:sat}). 
     
\section{Case $Z > 0$, Model II}\label{sec:modelII}

\begin{figure*}[t]
\includegraphics[width=\textwidth]{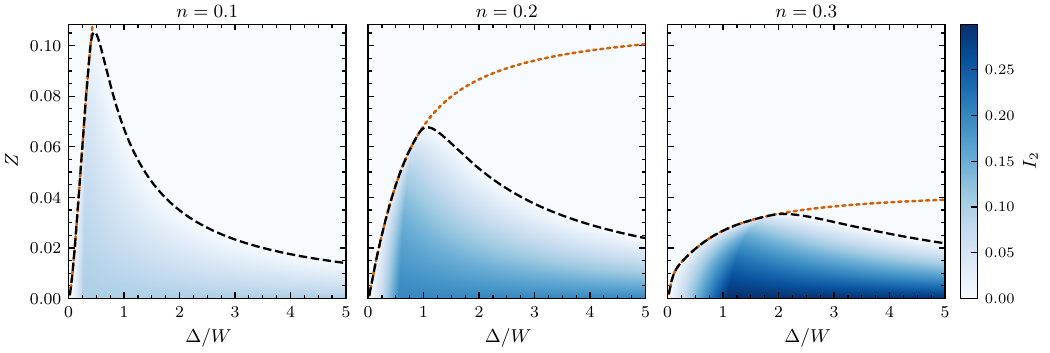}
\caption{\label{fig:I2mapII} 
The  Luttinger-Ward integral 
$I_{2}$ in Model II  [Eq.~(\ref{eq:VII:I2})]  for different $Z$ and $\Delta/W$ 
 for three filling factors $n <1/2$.
  The dashed line is the upper edge of the region in which $I_{2}\neq 0$. The dotted line is the condition (\ref{eq:VII:threshold}), beyond which the two zeros move away from the real axis. The two curves coincide at small $\Delta$, but differ at larger $\Delta$. The correct boundary $Z_c (\Delta)$, above which $I_2=0$, is the dashed line. At small $Z$, $I_{2}$ gradually deviates from its non-zero value at $Z=0$, Eq.~(\ref{eq:V:I1I2}).
}
\end{figure*}

\begin{figure}[t]
\includegraphics[width=\columnwidth]{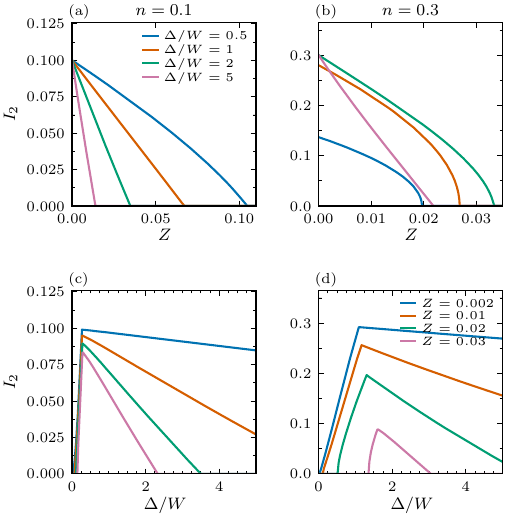}
\caption{\label{fig:I2cutsII} Cuts through $I_{2}(\Delta, Z)$ for Model II, presented in Fig.~\ref{fig:I2mapII}, for two values of $n$. Panels  (a), (b): $I_{2}$ versus $Z$ at fixed $\Delta$; panels (c), (d):  $I_{2}$ versus $\Delta$ at fixed $Z$. 
The Luttinger-Ward integral $I_{2}$ decreases monotonically with $Z$ and vanishes at 
 a critical $Z_c (\Delta)$ (the dashed line in Fig.~\ref{fig:I2mapII});  at $Z\to 0$,
 it approaches the value $n-\mathcal{V}(\mu_{\Delta })$, see Eq.~(\ref{eq:V:I1I2}). 
  As a function of $\Delta$, $I_2$  rises steeply at small $\Delta$, passes through a maximum, and then falls off as $\Delta$ becomes large enough such that $\mu_{\Delta }$ moves out of the band.}
\end{figure}

 In Model II,  the pole of $G_{FL} (\omega, \mathbf{k})$ is 
 at ${\bar \omega}=\xi_{\mathbf{k}}$ independent of $Z$.
 Proceeding as above,  we write the full Green's function as
 \begin{eqnarray}
&&G (\omega,\mathbf{k}) = 
\label{eq:VII:G}\\
&&
\frac{{\bar \omega}^{2} + (1-Z)\left(\eta_{\mathbf{k}} - \xi_{\mathbf{k}}\right) {\bar \omega} - \left[Z E^{2}_{\mathbf{k}} + (1-Z)\,\xi_{\mathbf{k}}\eta_{\mathbf{k}}\right]}{\left({\bar \omega} - \xi_{\mathbf{k}}\right)\left({\bar \omega}^2-E^2_{\mathbf{k}}\right)}
.
\nonumber
\end{eqnarray}
The full Green's function is again rational, with three poles and two zeros.
The pole count is the same as in Model I and yields 
$\mathcal{P}_{\mathbf{k}} = 1 + \Theta (-\xi_{\mathbf{k}})$, because $\xi_{\mathbf{k}}$ and $Z\xi_{\mathbf{k}}$ have the same sign. The zeros at ${\bar \omega} = \omega^{\pm}_{\mathbf{k}}$,  however, behave differently. Using $\eta_{\mathbf{k}} - \xi_{\mathbf{k}} = \mu_{FL} - \mu_{\Delta}$, which does not depend on $\mathbf{k}$, we find 
\begin{eqnarray}
&& {\omega}^{+}_{\mathbf{k}} {\omega}^{-}_{\mathbf{k}} = - F_{\mathbf{k}},~~
 {\omega}^{+}_{\mathbf{k}} + {\omega}^{-}_{\mathbf{k}} = 
(1-Z)(\mu_{\Delta} - \mu_{FL})
\nonumber \\
&& 
F_{\mathbf{k}} = \eta^{2}_{\mathbf{k}} - (1-Z)(\mu_{FL}-\mu_{\Delta})\,\eta_{\mathbf{k}} + Z\Delta^{2}.
\label{eq:VII:F}
\end{eqnarray}
In contrast to (\ref{eq:VI:prod}), $F_{\mathbf{k}}$ in (\ref{eq:VII:F}) is not positive definite.
 If $F_{\mathbf{k}}>0$, the two zeros have opposite signs,  as in Model I, and 
 $\mathcal{Z}_{\mathbf{k}} = 1$. If  $F_{\mathbf{k}}<0$, they have the same sign, fixed by 
  the value of their sum.  For $n <1/2$, $\mu_\Delta < \mu_{FL}$ (both are negative), hence    
  $\omega^{+}$ and $\omega^{-}$  are both  negative and 
  $\mathcal{Z}_{\mathbf{k}} =2$.  
  
 Combining these results and 
 integrating over the BZ, we obtain 
\begin{equation}
I_1 = n - \int_{F_{\mathbf{k}}<0} \frac{d^{2}k}{(2\pi)^{2}},~~ I_2 = \int_{F_{\mathbf{k}}<0} \frac{d^{2}k}{(2\pi)^{2}}.
\label{eq:VII:I2}
\end{equation}
  Because $F_{\mathbf{k}}$ is quadratic in $\epsilon_{\mathbf{k}}$, the region $F_{\mathbf{k}}<0$ is the strip $\bar{\epsilon} - R < \epsilon_{\mathbf{k}} < \bar{\epsilon} + R$ between two contours, with
\begin{eqnarray}
&& \bar{\epsilon} = \frac{(1+Z)\mu_{\Delta} + (1-Z)\mu_{FL}}{2}
\nonumber \\
&& R = \sqrt{\left(\frac{(1-Z)(\mu_{FL}-\mu_{\Delta})}{2}\right)^{2} - Z\Delta^{2}}
.
\label{eq:VII:region}
\end{eqnarray}
The strip exists only when $R$ is real, i.e., when 
\begin{equation}
(1-Z)^{2}\left(\mu_{FL}-\mu_{\Delta}\right)^{2} > 4 Z \Delta^{2}
.
\label{eq:VII:threshold}
\end{equation}
If this condition does not hold,  there is no range where $F_{\mathbf{k}} <0$. Then $I_1 =n$ and $I_2 =0$.
Because  neither $\mu_{FL}$ nor $\mu_{\Delta}$ depends on $Z$, 
 $I_2$ decreases with $Z$ but remains non-zero  until $Z$ reaches some critical value $Z_c (n, \Delta)$.
 This  behavior is  qualitatively different from that of Model I, where $I_2$ changes discontinuously.   On a more careful look, we find that $Z_c (n, \Delta)$, determined by 
  Eq.~(\ref{eq:VII:threshold}), sets the boundary above which 
  $I_{2}$  vanishes, only for $\Delta \leq W$. At larger $\Delta$,  when $|\mu_{\Delta }|$ exceeds $W/2$,   $Z_c$ shrinks to a smaller value as $I_2$ vanishes, even when the strip specified by (\ref{eq:VII:region}) exists but 
  is located  outside the bandwidth of $\epsilon_{\mathbf{k}}$. We show the behavior of $I_{2}(\Delta ,Z)$  in Fig.~\ref{fig:I2mapII} and show representative cuts in 
Fig.~\ref{fig:I2cutsII}.  

 Returning to the smallest $Z$, we see that  in  Model II, $I_{1,2}$ displays continuous behavior, whereas in Model I, $I_{1,2}$ jumps at $Z =0+$. The reason is the position of the pole in the Fermi-liquid component of the Green's function: in Model II, it remains at ${\bar \omega}=\xi_{\mathbf{k}}$ as $Z \to 0$ and is canceled in $I_1$ by a Green's function  zero approaching the same point,  while in Model I there is a range of $\mathbf{k}$ where 
 the pole  and the zero approach the origin from different sides of the real axis, and only one of the two contributes to $I_1$.

\section{Particle-hole symmetry}\label{sec:PH}

\begin{figure}[t]
\includegraphics[width=\columnwidth]{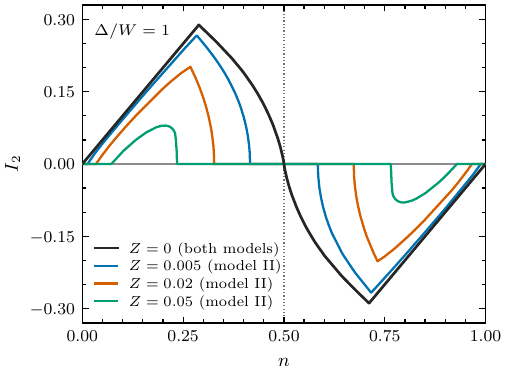}
\caption{\label{fig:I2vsN} The Luttinger-Ward integral $I_{2}$ as a function of filling $n$ 
for  $\Delta/W=1$ and different $Z$.  Every curve is odd about half-filling and vanishes at $n=1/2$. The black curve is the $Z=0$ result (\ref{eq:V:I1I2}), the colored curves are for Model II at three values of $Z$.  For each non-zero $Z$,  there is a finite window around half-filling where $I_{2} =0$; the window widens with increasing $Z$.   In Model I, $I_{2}$ vanishes identically for all $Z>0$.
}
\end{figure}

In this section, we compare the forms of the Luttinger-Ward integrals $I_1$ and $I_2$ at $0<n<1/2$ and $1/2<n<1$.  
 The tight-binding dispersion of Sec.~\ref{sec:model} satisfies $\epsilon_{\mathbf{k}+\mathbf{Q}} = -\epsilon_{\mathbf{k}}$ with $\mathbf{Q} = (\pi,\pi)$. The  density of states is  then
  even under $\mathbf{k} \to \mathbf{k}+\mathbf{Q}$, and $\mu_\Delta$ and $\mu_{FL}$ are both odd under $n \to 1-n$.  This implies that right at half-filling ($n=1/2$), $\mu_\Delta = \mu_{FL} =0$.  The Green's function for both models $G (\omega,\mathbf{k},n)$ then satisfies 
\begin{equation}
G (\omega,\mathbf{k},n) = G (-\omega,\mathbf{k}+\mathbf{Q}, 1-n).
\label{eq:PHG}
\end{equation}
  Every pole and every zero of $G$ is thus reflected about the origin,
   so that those above the real axis are exchanged with those below the real axis.
  An elementary analysis then shows that under the substitution $n \to 1-n$, the two Luttinger-Ward integrals $I_{1} \equiv I_{1,n}$ and $I_2 \equiv I_{2,n}$ become 
\begin{equation}
I_{1, 1-n} = 1 - I_{1,n}
,\qquad 
I_{2,1-n} = - I_{2,n}
.
\label{eq:PHI2}
\end{equation}
We emphasize that (\ref{eq:PHI2}) holds for both models and for any $Z$ and $\Delta$; it relies only on the condition $\epsilon_{\mathbf{k}+\mathbf{Q}} = -\epsilon_{\mathbf{k}}$. These relations do not hold in the presence of next-nearest-neighbor hopping. We illustrate (\ref{eq:PHI2}) in Fig.~\ref{fig:I2vsN}, where every curve is odd about $n=1/2$.

At half-filling, $n=1/2$,  the transformation maps the model onto itself. Equations (\ref{eq:PHI2}) then become  $I_{1, 1/2} = 1 - I_{1,1/2}$ and  $I_{2,1/2} = -I_{2,1/2}$, yielding
\begin{equation}
I_{2} = 0
,\qquad 
I_{1} = \frac{1}{2} =n,
\label{eq:halffilling}
\end{equation}
again for both models and for any $Z$ and $\Delta$, including $Z=0$. We emphasize that 
this is the outcome of the symmetry analysis alone -- there is no need to compare the poles and zeros of the Green's function. 

\section{Conclusions}

In this paper, we addressed the issue of whether a pole in the fermionic self-energy 
 invalidates the Luttinger-Ward analysis of the two integrals $I_1$ and $I_2$, which were used to derive the relation between the Fermi surface area and the fermionic density, known as the Luttinger theorem. Our short answer is: no.  We identified two models
 in which the self-energy  has a pole, yet  the Luttinger-Ward relations $I_1 = n$, $I_2 =0$, where $n$ is the fermionic density, hold.  In both models, a self-energy with a pole transforms a  portion of the spectral weight into a gapped continuum, but some weight remains in the residual Fermi liquid component. 
 
 We used a checkerboard polaron state of spinless fermions interacting with an Einstein phonon as a model for a gapped continuum coming from the self-energy with a pole. For this model alone, Luttinger-Ward reasoning does not hold, and $I_2$ is non-zero and increases with the residue of the pole in the self-energy (our parameter $\Delta^2$). 
 
 We considered two models for the residual Fermi liquid, Model I and Model II. In both models, the residual Fermi liquid gives rise to a non-zero spectral weight around $\omega =0$ with the 
   residue $Z \ll 1$. In 
 Model I, the same small $Z$ gives rise to the renormalization of the quasiparticle bandwidth $W$ into $WZ$.  In Model II,  the bandwidth is not renormalized by $Z$. 

For Model I,  we obtain $I_{2}=0$ and $I_{1}=\mathcal{V}(\mu_{FL})=n$ for {\em any} $Z>0$, Eq.~(\ref{eq:VI:I2}). In this situation,  Luttinger-Ward reasoning holds in its original form even when $Z$ is arbitrarily small, despite the fact that most of the spectral weight is in the two gapped continua and the density of states vanishes identically between the renormalized quasiparticle bandwidth and the onset of the continuum, Fig.~\ref{fig:DOS}(a). We argued that the discontinuity of the Luttinger-Ward integrals at $Z = 0+$ is the consequence of the singularity of the Fermi liquid component of the Green's function at $Z \to 0$ and $\omega \to 0$, which makes the order of taking the limits relevant.  

For  Model II, we find that the Luttinger-Ward integrals evolve continuously with $Z$, and there exists a finite critical $Z_c (n, \Delta)$, above which the original 
Luttinger-Ward reasoning holds.

For both models, we used tight-binding dispersion, which obeys particle-hole symmetry. For such a dispersion, we argued that right at half-filling, $I_2 =0$ at any $Z$, including $Z=0$
 just by symmetry arguments. 

   \begin{acknowledgments}
We acknowledge with thanks useful discussions with E. Berg, I. Esterlis, M. Fabrizio, L. Glazman, N. Perkins, Ph. Phillips, D. Maslov, N. Prokof'ev,  S. Sachdev, G. Sangiovanni, and J. Schmalian. 
The 
figures
were done
with the assistance of Claude (Anthropic).
AVC was supported by the U.S.\ Department of Energy, Office of Science, Basic Energy Sciences, under Award No.~DE-SC0014402. ArA was supported by the DOE Office of Science, DE-SC0026038.
\end{acknowledgments}

\bibliography{luttinger}

   \end{document}